\documentclass[preprint,aps,prl,amsmath,amssymb,superscriptaddress,notitlepage]{revtex4}
\usepackage{graphicx}
\usepackage{color}
\usepackage{bm}
\usepackage{float}
\usepackage{amsmath,amssymb}

\begin{document}

\bibliographystyle{apsrev}

\title{Coexistence of Charge- and Ferromagnetic-Order in fcc Fe}

\author{Pin-Jui Hsu}
\email[corresponding author: ]{phsu@physnet.uni-hamburg.de}
\affiliation{Physikalisches Institut, Experimentelle Physik II, 
		Universit\"{a}t W\"{u}rzburg, Am Hubland, D-97074 W\"{u}rzburg, Germany}
\author{Jens K{\"u}gel}
\affiliation{Physikalisches Institut, Experimentelle Physik II, 
		Universit\"{a}t W\"{u}rzburg, Am Hubland, D-97074 W\"{u}rzburg, Germany}
\author{Jeannette Kemmer}
\affiliation{Physikalisches Institut, Experimentelle Physik II, 
		Universit\"{a}t W\"{u}rzburg, Am Hubland, D-97074 W\"{u}rzburg, Germany}
\author{Francesco Parisen Toldin}
\address{Institut f{\"u}r Theoretische Physik und Astrophysik, 
	Universit{\"a}t W{\"u}rzburg, 97074 W{\"u}rzburg, Germany,}
\author{Tobias Mauerer}
\affiliation{Physikalisches Institut, Experimentelle Physik II, 
		Universit\"{a}t W\"{u}rzburg, Am Hubland, D-97074 W\"{u}rzburg, Germany}
\author{Matthias Vogt}
\affiliation{Physikalisches Institut, Experimentelle Physik II, 
		Universit\"{a}t W\"{u}rzburg, Am Hubland, D-97074 W\"{u}rzburg, Germany}
\author{Fakher Assaad}
\address{Institut f{\"u}r Theoretische Physik und Astrophysik, 
	Universit{\"a}t W{\"u}rzburg, 97074 W{\"u}rzburg, Germany,}
\author{Matthias Bode} 
\affiliation{Physikalisches Institut, Experimentelle Physik II, 
		Universit\"{a}t W\"{u}rzburg, Am Hubland, D-97074 W\"{u}rzburg, Germany}
\affiliation{Wilhelm Conrad R{\"o}ntgen-Center for Complex Material Systems (RCCM), 
		Universit\"{a}t W\"{u}rzburg, Am Hubland, D-97074 W\"{u}rzburg, Germany}

\date{\today}
\maketitle

{\bf 
Phase coexistence phenomena have been intensively studied in strongly correlated materials where several ordered states simultaneously occur or compete. 
Material properties critically depend on external parameters and boundary conditions, 
where tiny changes result in qualitatively different ground states.  
However, up to date, phase coexistence phenomena have exclusively been reported 
for complex compounds composed of multiple elements. 
Here we show that charge- and magnetically ordered states coexist in double-layer Fe/Rh(001). 
Scanning tunneling microscopy and spectroscopy measurements 
reveal periodic charge-order stripes below $T_{\rm P} = 130$\,K. 
At $T = 6$\,K they are superimposed by ferromagnetic domains as observed by spin-polarized scanning tunneling microscopy.  
Temperature-dependent measurements reveal a pronounced cross-talk 
between charge- and spin-order at the ferromagnetic ordering temperature $T_{\rm C} \approx 70$\,K, 
which is successfully modeled within an effective Ginzburg-Landau ansatz including sixth-order terms.  
Our results show that subtle balance between structural modifications 
can lead to competing ordering phenomena.
}

\pagebreak
In the recent past competing order phenomena, such as the interplay between spin- and charge-order 
in copper- and iron-based superconductors \cite{JHoffman, JDavis, PCai}, 
the magnetic modulation--induced emergence of spontaneous polarization 
in multiferroics \cite{PRL85_3720,TKimura, MKenzelmann, SCheong}, 
or the coexistence of magnetism and superconductivity at the interface 
of oxide heterostructures \cite{DADikin, LLi, JABert} have intensively been investigated.  
In these materials subtle changes of the chemical composition or external stimuli 
may eventually lead to nontrivial emergent excitations at quantum critical points 
between lowest energy states \cite{Sachdev_Science}.

Although iron (Fe) is usually considered the prototypical ferromagnetic material, 
it exhibits strong correlations between atomic, orbital, and magnetic spin structure, 
that result in a rich variety of interesting magnetic properties. 
Bulk Fe crystallizes in a body-center-cubic (bcc) crystal structure
and shows robust ferromagnetism (FM) with a Curie temperature $T_{\rm C} = 1043$\,K~\cite{Kittel}. 
For low-dimensional Fe ultra-thin films and nanostructures, however, various magnetic ground states 
and nontrivial spin textures have been theoretically predicted, including non-magnetic~\cite{CSWang}, 
non-collinear antiferromagnetic (AFM) ordering~\cite{FJPinski, TAsada}, 
incommensurate spin-density wave (SDW)~\cite{DSpisak}, 
helical spin spiral (SS)~\cite{KKnopfle}, and magnetic skyrmions~\cite{SHeinze}. 

Recent advanced experimental studies~\cite{DQian, AKubetzka, SMeckler, NRomming} indicate, 
that these apparently contradicting reports are caused by the fact that the magnetic ground state of Fe 
is highly sensitive to the interplay of electronic hybridization and structural instabilities in reduced dimensions.
In particular, the magnetism of ultra-thin pseudomorphic Fe films on fcc Rh(001), 
which has been subject of several investigations~\cite{PhysRevB.64.054417,Hayashi_JPSJap,%
DSpisak2,AlZubi,MTakada,KWK2015}, appears to be strongly influenced by the competition 
between ferromagnetic order in bcc $\alpha$- and antiferromagnetism in fcc $\gamma$-Fe 
as well as electronic hybridization of the film's $3d$ with substrate's $4d$ states \cite{DSpisak2}.
While the monolayer exhibits an antiferromagnetic c$(2 \times 2)$ spin structure%
~\cite{DSpisak2,AlZubi,KWK2015}, films with a local thickness of 2 and 3\,atomic layers (AL) are ferromagnetically ordered 
with the easy axis of magnetization along the surface normal~\cite{MTakada,KWK2015}.
It has been speculated that the competition between antiferromagnetic and ferromagnetic order 
in tetragonally distorted films may lead to low-energy excitations or competing phase transitions \cite{DSpisak2}.

In this study, we report on the observation of a phase coexistence phenomenon 
in a pseudomorphic Fe double-layer films grown on Rh(001).  
Our scanning tunneling microscopy (STM) and spectroscopy (STS) measurements 
reveal that different order phenomena, i.e.\ charge- and ferromagnetic spin-order, 
coexist at low temperatures for below the respective phase transitions.  
Interestingly, we observe observe a remitted reduction of the charge-order parameter $\phi$ 
at the ferromagnetic Curie temperature, indicating a crosstalk between charge- and spin-order.
This behavior is successfully be modeled by Ginzburg-Landau (GL) calculations. 
We speculate that this crosstalk may be mediated by electronic states at or very close to the Fermi level. 
Since the system investigated here is structurally much more simple than other materials with coexisting order phenomena, 
it may become a model system and allow for a better understandings of competing order phenomena.

\section{Results}
\begin{figure}[t]
	\includegraphics[width=0.99\columnwidth]{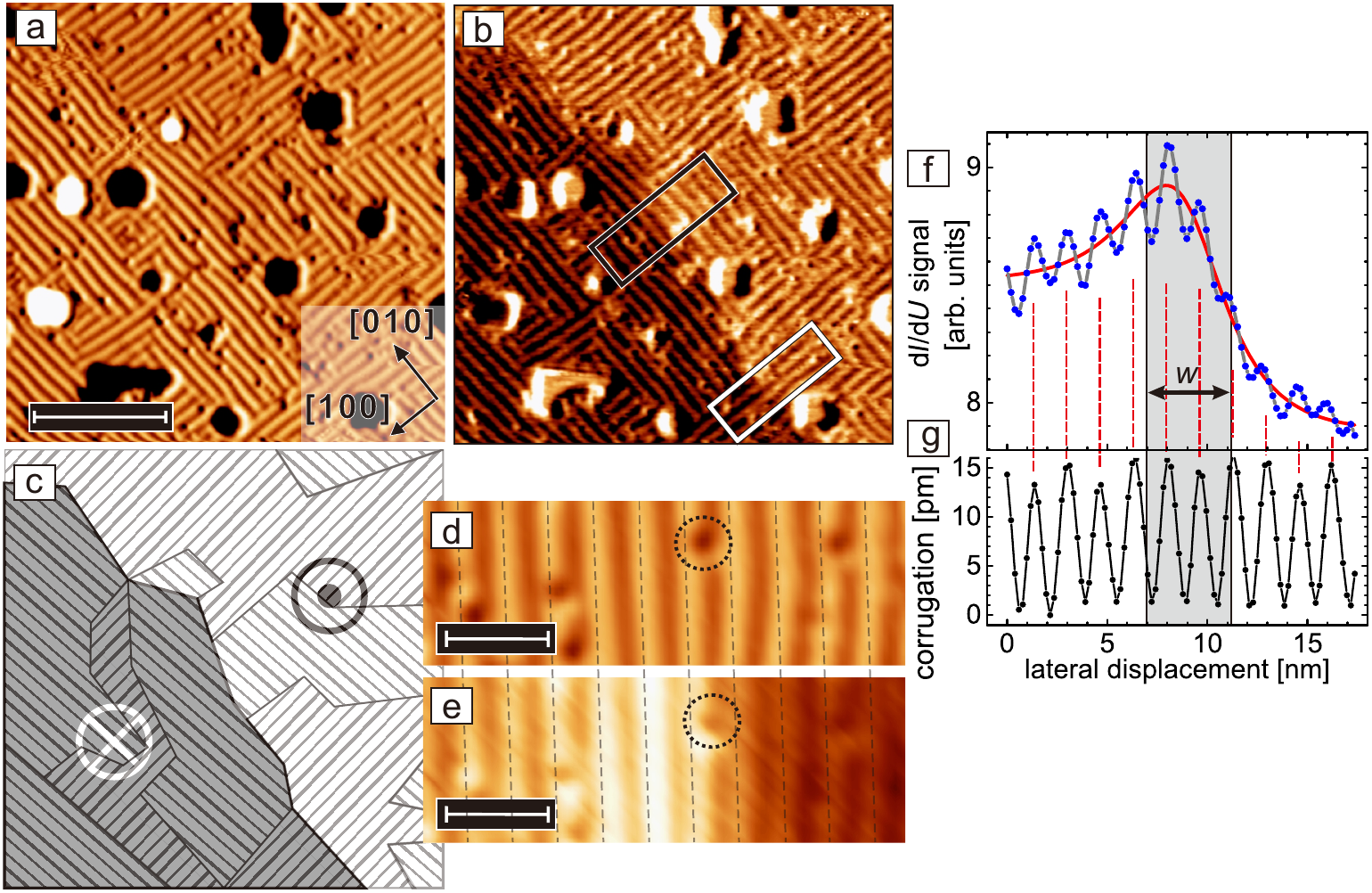}
	\caption{{\bf Coexistence of magnetic domains and a striped phase.} 
	({\bf a})~Topography and ({\bf b})~the simultaneously measured spin-resolved $\mathrm{d}I/\mathrm{d}U$ map 
	of $(1.95 \pm 0.02)$\,AL Fe/Rh(001) showing out-of-plane magnetic domains 
	(scan parameters: $U = -0.7$\,V, $I = 500$\,pA, $T = 5$\,K). Scale bar is 15 nm.  
	The white and black boxes mark regions where the stripe superstructure 
	is oriented parallel (${\mathbf{q}_{1}}$) or perpendicular (${\mathbf{q}_{2}}$) 
	to the magnetic domain wall, respectively.
	({\bf c}) Schematic drawing of the magnetic domain and stripe patterns observed in {\bf b}.
	({\bf d}),({\bf e}) Zoomed-in topographic and $\mathrm{d}I/\mathrm{d}U$ image of a location 
	similar to the one shown in the black box in {\bf b} ($U = -1.0$\,V, $I = 500$\,pA). Scale bars are 3 nm.
	({\bf f}),({\bf g}) Line sections obtained from {\bf e} and {\bf d}, 
	showing spin-resolved $\mathrm{d}I/\mathrm{d}U$ signal together with topographic corrugation.
	Fitting the domain wall (red line) gives a wall width $w = (4.32 \pm 0.35)$\,nm.
	Note, that the presence of the domain wall does not significantly influence 
	the modulation of the stripe superstructure, as indicated by the equally spaced dashed red lines. 
	\label{fig:Magn_and_stripes} }
\end{figure}

\textbf{Coexistence of charge- and ferromagnetic spin-order.} Figures\,\ref{fig:Magn_and_stripes}a and b show the topography 
and the differential conductivity $\mathrm{d}I/\mathrm{d}U$, respectively, 
of a $(1.95 \pm 0.02)$\,AL Fe film on Rh(001) resolved by using spin-polarized scanning tunneling microscopy (SP-STM) at $T = 5$\,K.   
An almost perfectly closed double-layer is obtained, with a few holes  
and some tiny triple-layer islands as the only imperfections.
Both data sets were measured simultaneously using a Cr-coated probe tip 
with out-of-plane magnetic sensitivity~\cite{MBode1}. 
The magnetic contrast is particularly well visible in Fig.\,\ref{fig:Magn_and_stripes}b
which shows the tunneling magneto-resistance (TMR) contrast 
of two domains with opposite perpendicular magnetization 
as signaled by the dark and bright regions in the lower left and upper right of the image.  

The Curie temperature $T_{\rm C}$ of this double-layer turns out to be very low. 
While $T_{\rm C}$ of a 3.0\,AL Fe film on Rh(001) amounts to approximately 320\,K,
a steep linear decrease has been observed towards thinner films \cite{Hayashi_JPSJap}.  
For the ferromagnetic double-layer no magnetic signal 
could be detected down to $T = 97$\,K \cite{Hayashi_JPSJap}.
Linear extrapolation to an Fe coverage of 2.0\,AL \cite{Hayashi_JPSJap} 
suggests a Curie temperature below 80\,K. 
In fact, our temperature-dependent SP-STM measurements confirm this finding (see Supplementary Figure 1).   
It has been speculated that the surprisingly low Curie temperature may be related 
to the above-mentioned FM--AFM competition which potentially results in a very low Curie temperature 
and/or excited magnetic states at relatively low excitation energies \cite{DSpisak2}.  
As we will show below the situation appears to be even more complex, 
with different competing ordering phenomena at work. 

Interestingly, Fig.\,\ref{fig:Magn_and_stripes}a and b also reveal that the out-of-plane FM spin-order 
of the Fe double-layer on Rh(001) coexists with a periodic one-dimensional superstructure 
that consists of stripes along the [100] and [010] directions of the substrate. 
For clarity the coexisting magnetic domain and stripe patterns observed in Fig.\,\ref{fig:Magn_and_stripes}b
is schematically represented by dark/bright background and differently oriented periodic lines 
in Fig.\,\ref{fig:Magn_and_stripes}c, respectively.
The periodicity of the stripes amounts to $(1.48 \pm 0.22)$\,nm, corresponding to 
a superstructures with wave vectors ${\mathbf{q}_{1}} = (2 \pi / a)(0.26 \pm 0.03,0,0)$ 
and $\mathbf{q_{2}} = (2 \pi / a)(0,0.25 \pm 0.03,0)$, 
where $a$ is the Rh(001) atomic lattice constant of 3.80{\AA}. 

Detailed analysis indicates that at the measurement temperature of 5\,K 
there is no significant correlation between the stripe pattern and the magnetic domain structure. 
For example, the black and white boxes in Fig.\,\ref{fig:Magn_and_stripes}b 
mark surface areas where the stripe superstructure is oriented parallel (${\mathbf{q}_{2}}$) 
or perpendicular (${\mathbf{q}_{1}}$) to this domain wall, respectively.  
In neither case the presence of the domain wall seems to have any significant influence on the stripes.

This impression is also confirmed by the analysis of an area 
with a configuration similar to the black box in Fig.\,\ref{fig:Magn_and_stripes}b, 
i.e.\ with stripes along the [010] directions (${\mathbf{q}_{2}}$) across a magnetic domain wall. 
Fig.\,\ref{fig:Magn_and_stripes}d and e show the topography 
and the $\mathrm{d}I/\mathrm{d}U$ map, respectively.
An individual defect is marked by circles in both images. 
Fig.\,\ref{fig:Magn_and_stripes}f presents line sections 
drawn along the long axes of these images.
Comparison of the peak position of the stripe superstructure 
in both the topographic as well as the spin-resolved $\mathrm{d}I/\mathrm{d}U$ channel 
shows no indication for any changes across the domain wall 
(see dashed red vertical lines in Fig.\,\ref{fig:Magn_and_stripes}e).

\begin{figure}[t]
	\includegraphics[width=0.99\columnwidth]{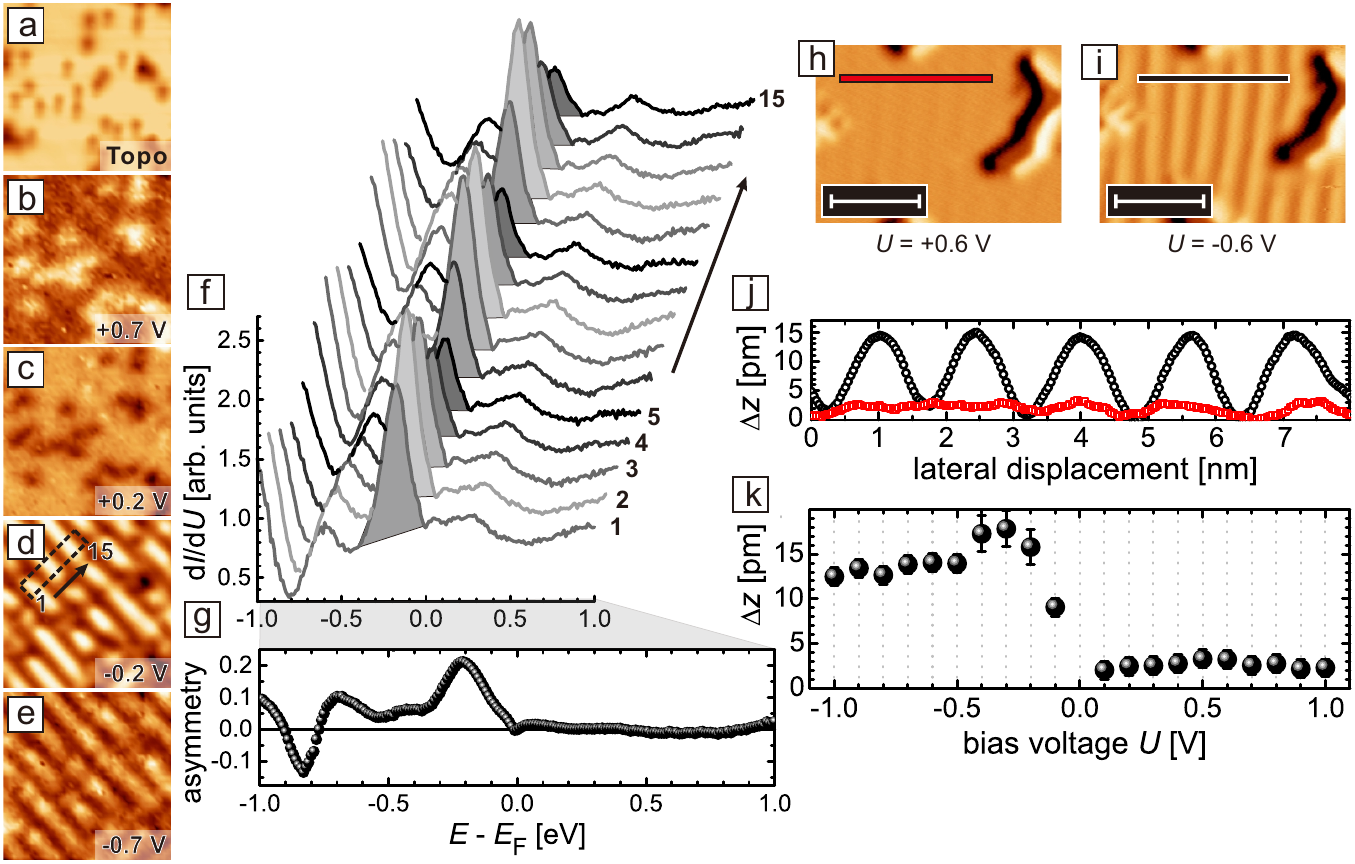}
	\caption{{\bf Scanning tunneling spectroscopy and bias voltage--dependent corrugation.} 
	({\bf a})~Topography of the Fe double-layer on Rh(001) and 
	({\bf b})-({\bf e})~d\textit{I}/d\textit{U} maps taken at the indicated bias voltages 
	from constant-separation STS data (setpoint parameters: $U = +1.0$\,V, $I = 500$\,pA). 
	The stripe pattern is only visible in the occupied energy range. Image sizes are 15 x 15 nm$^{2}$.
	({\bf f})~Tunneling spectra measured along the box in {\bf d}. 
	The peak at $-0.2$\,V is more intensive on the bright stripes than between them.  
	({\bf g})~Plot of the electronic asymmetry as a function of bias voltages.  
	While the electronic asymmetry is negligible in the empty states, 
	it becomes maximal in the occupied states at about $-0.2$\,V.   
	({\bf h}),({\bf i})~Topographic images taken at $U = +0.6$\,V and $-0.6$\,V. Scale bars are 5 nm. 
	({\bf j}) Averaged line sections measured along the red and black line.
	({\bf k})~Bias-dependence of the corrugation. 
	Error bars are given by standard deviation of corrugation peak heights.    
	\label{fig:Spec} }
\end{figure}
\textbf{Electronic structure of charge-ordered phase.} 
In order to unravel the physical origin of the stripes we have performed STS measurements 
to probe the local density-of-states (LDOS) of the region shown in Fig.\,\ref{fig:Spec}a (topography). 
Fig.\,\ref{fig:Spec}b-e shows a series of $\mathrm{d}I/\mathrm{d}U$ maps 
extracted from this data set at some representative bias voltages $U$. 
Obviously, the appearance of stripes strongly depends on $U$.  
While the stripes cannot be detected within the signal-to-noise ratio 
at positive bias voltages, i.e.\ when tunneling into empty sample states, 
they are clearly visible at negative bias (occupied states).  
As can be seen in curves 1 through 15 of Fig.\,\ref{fig:Spec}f 
(obtained within the box in Fig.\,\ref{fig:Spec}d), 
the spectra are characterized by a pronounced peak at $-0.2$\,V,
a weaker peak at $-0.6$\,V, and a dip at $-0.8$\,V (which are all absent 
in the relatively featureless spectrum of the Fe monolayer; not shown here). 
The sequence of spectra reveals that the peak intensity varies periodically.   
Obviously, the main peak at $-0.2$\,V appears much more intense 
when the tip positioned above a bright stripe (see, e.g., spectrum 2) 
than above a dark one (spectrum 5).  

The sign and intensity of the bias-dependent contrast of the striped superstructure 
can be analyzed more systematically by calculating the energy-dependent asymmetry
which is defined as the difference of the differential conductance 
measured on and off a bright stripe at a particular energy, $E - E_{\rm F}$, 
divided by their sum, i.e.\ 
${\left( \mathrm{d}I/\mathrm{d}U \right)_{\rm on}} 
	\mathbin{/} 
	{\left( \mathrm{d}I/\mathrm{d}U \right)_{\rm off}}$. 
The asymmetry curve calculated from tunneling spectra is plotted in Fig.\,\ref{fig:Spec}g. 
While the asymmetry is negligible at positive sample bias 
several features can be recognized at negative bias voltages, 
with a pronounced asymmetry maximum at $-0.2$\,V.

Fig.\,\ref{fig:Spec}h and i exemplarily show two topographic STM images 
obtained at the same sample location.  
While only subtle modulations can be recognized at $U = +0.6$\,V (Fig.\,\ref{fig:Spec}h)
the stripes are clearly resolved at $U = -0.6$\,V (Fig.\,\ref{fig:Spec}i). 
The corresponding line profiles in Fig.\,\ref{fig:Spec}j reveal a corrugation of about 14\,pm at $U = -0.6$\,V, 
whereas it is below $2$\,pm at $U = +0.6$\,V.
Fig.\,\ref{fig:Spec}k summarizes the observed bias-dependence of the corrugation. 
Apparently the corrugation is extremely low at positive bias,  
rises up to a maximum value of about 17\,pm at $U \approx -0.2 ... 0.4$\,V---a value 
that is in line with the above-mentioned energy-dependent asymmetry 
(cf.\ Fig.\,\ref{fig:Spec}g)---and then slowly decreases to $\approx 12$\,pm at $U = -1$\,V.  

\begin{figure}[t]
    \includegraphics[width=0.7\columnwidth]{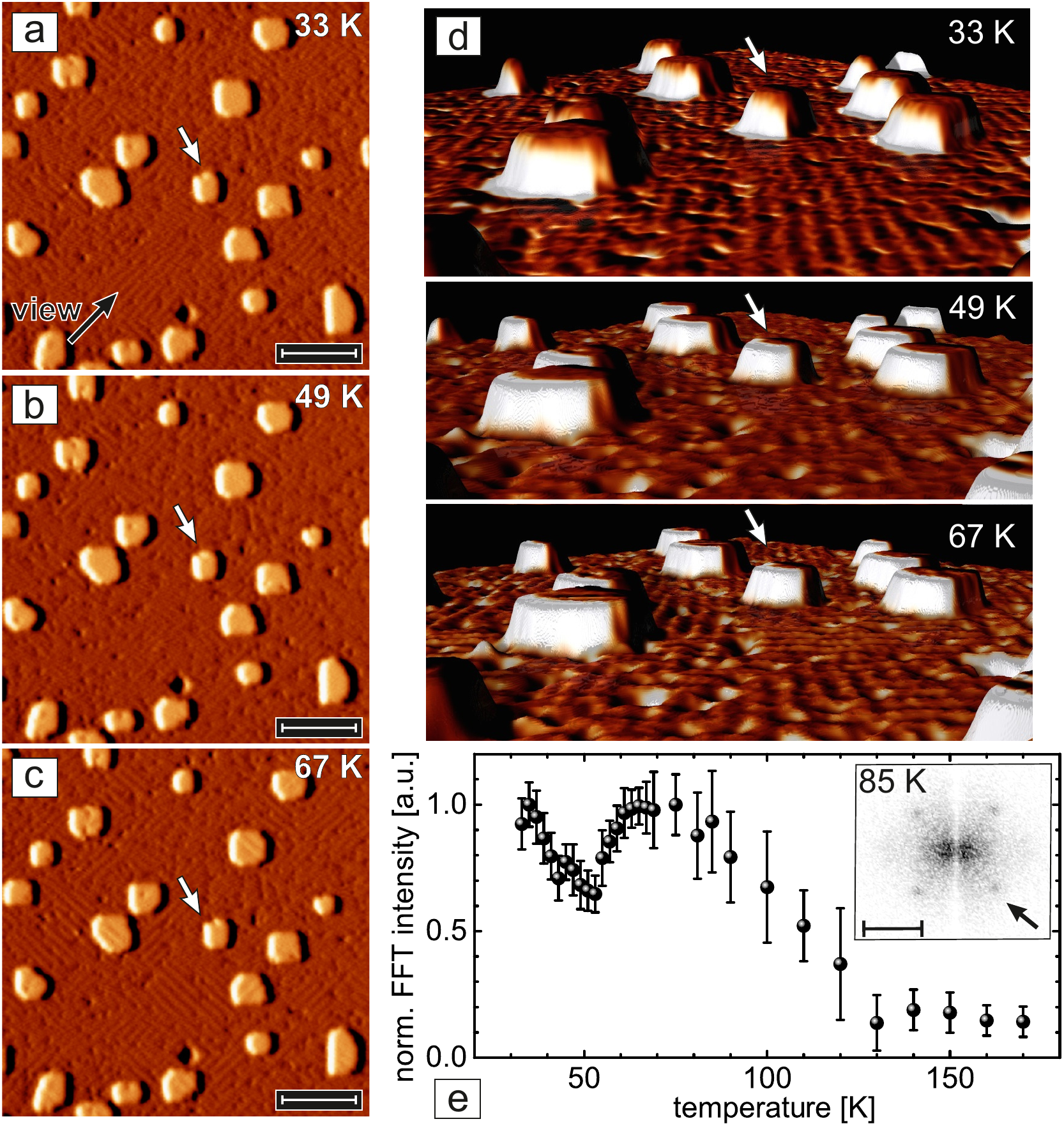}
	\caption{{\bf Temperature-dependent charge ordering.} 
	({\bf a})-({\bf c}) STM topographic showing the very same location 
	of the sample surface at $T = 33$\,K, $49$\,K, and $67$\,K. 
	The scale bars are 20\,nm long.  One island is marked by white arrows.  
	({\bf d}) Rendered perspective images as seen along the viewing direction 
	marked by a black arrow in (a).  
	While the stripes are clearly visible at low and high temperatures 
	(see, e.g., the area in front of the island marked by a white arrow), 
	the intensity is strongly reduced at the intermediate temperature. 
	({\bf e})~Plot of the temperature-dependent normalized intensity 
	of the spots indicative for the stripe pattern (see arrow).
	A dip starting at around $T = 70$\,K can be recognized.  
	An example of a Fourier-transformed STM image taken at $T = 85$\,K is shown in the inset. 
	Error bars represent the spot's full width at hals maximum after subtraction of the background intensity. 
	Scale bar is 0.5\,nm$^{-1}$.
	\label{fig:VT} }  
\end{figure}

{\bf Temperature-dependent electronic reconstruction.} 
While the measurements presented so far suggest the existence of two apparently independent 
ordering phenomena, i.e.\ ferromagnetism and the formation of stripes with a pronounced LDOS modulation, 
the following data indicate that the two phenomena are coupled and influence each other.  
Fig.\,\ref{fig:VT}a-c show three STM topographic images of a 2~AL Fe film on Rh(001) taken at $T = 33$\,K, 49\,K, and 67\,K.
In order to exclude any potential influence of local fluctuations of sample quality 
all data were taken at the same location as emphasized by white arrows pointing at one particular island.
Since the rather small corrugation of the stripes indicative for charge-order is difficult to recognize, 
the corresponding rendered perspective images of Fig.\,\ref{fig:VT}a-c 
(viewing direction indicated by a black arrow) are displayed in Fig.~\ref{fig:VT}d.  
While stripes are clearly visible at 33\,K (Fig.\,\ref{fig:VT}a), 
a much lower corrugation amplitude can be recognized at 49\,K (Fig.\,\ref{fig:VT}b).  
Surprisingly, the intensity of the stripes increases again 
as the temperature is further increased to $T = 67$\,K (Fig.\,\ref{fig:VT}c). 

Figure\,\ref{fig:VT}e presents a summary of several temperature-dependent data sets 
that were obtained through fast Fourier transformation (FFT) of constant-current STM images. 
As can be seen in the inset of Fig.\,\ref{fig:VT}e this results in four spots---one of which 
is marked by an arrow---indicative of the above-mentioned superstructures with ${\mathbf{q}_{1,2}}$. 
Starting at the lowest temperature accessible with our variable-temperature STM, i.e.\ 35\,K,
the normalized intensity of these FFT spots first decreases 
with increasing temperature up to $T \approx 50$\,K. 
When increasing the temperature further, however, an unexpected upturn of the FFT intensity is observed 
until approximately the initial value is reached at $T \approx 70$\,K.  
Raising the temperature beyond this value leads to another reduction of the FFT intensity 
until it eventually becomes indistinguishable from the background 
above the charge-order transition temperature $T_{\rm P} = (128 \pm 12)$\,\,K. 
We note that this temperature dependence is continuous and reversible, 
i.e.\ the periodic modulations as well as the spots in the FFT image 
reappear as the temperature is lowered, consistent with a second-order phase transition
and excluding any potential aging or contamination effects. 

Our experimental observations indicate that there are two competing ordering phenomena at work:
(i) Charge-order sets in at about 130\,K and results in stripes visible at negative bias voltages and 
(ii) ferromagnetic order which can be observed by spin-polarized STM below about 75\,K. 
Since both ordering phenomena can be explained by Fermi surface instabilities, 
some degree of cross-correlation can readily be expected. 
It should be noted that in many cases the onset of charge-order coincides with electronic structure changes 
which are not largest directly at the Fermi level but at slightly different binding energies.  
For example, when cooling through the charge-density wave phase transition temperatures 
of 1$T$-TiSe$_2$ \cite{Claessen1990,Rossnagel2002} or 1$H$-TaSe$_2$  \cite{Valla2000}
in temperature-dependent photoemission spectroscopy experiments 
the strongest variation in the energy distribution curves was observed at $E - E_{\rm F} = -0.2$\,eV, 
i.e.\ at a binding energy similar to what we present in Fig.\,\ref{fig:Spec}(f).
We can only speculate why the peak that is indicative of charge-order in our STS spectra 
doesn't appear directly at but 200\,meV below the Fermi level.  
One possibility would be that the responsible bands somewhat disperse
and exhibit tunneling matrix elements that are higher for states 
which are still involved in the phase transition but further away from the Fermi level. 

{\bf Ginzburg-Landau theory.} Indeed, the observed phenomenology can be modeled within a Ginzburg-Landau (GL) theory. 
Its formulation is dictated by the universal properties of the system, 
such as the number of components of the order parameter and the symmetries of the system. 
In the present case there are two order parameters describing the charge and magnetic order. 
Pinning of the charge order and magnetic anisotropy allow to consider two scalar order parameters, 
the charge order parameter $\phi$, which can be identified with the intensity of the FFT spots, 
and the Ising-like magnetization $m$. 
The systems exhibits a ${\mathbb Z}_2$ symmetry on both order parameters, 
$\phi\rightarrow -\phi$, $m\rightarrow -m$, 
so that the global symmetry group is ${\mathbb Z}_2\oplus{\mathbb Z}_2$. 
A GL free energy is obtained by expanding the free energy $F$ 
in powers of $\phi$ and $m$, retaining only the terms 
which respect the given symmetry group (see, e.g., Ref.~\cite{critbook}):
\begin{multline}
	F = F_0 + \frac{a}{2}\left(T-T_{\rm P}\right)\phi^2 + \frac{b}{4}\phi^4
		+ \frac{a'}{2}\left(T-T_{\rm C'}\right)m^2+\frac{b'}4m^4+\frac{\gamma}{2}\phi^2m^2,
\label{LandauF}
\end{multline}
where $\gamma$ is the coupling constant between $\phi$ and $m$, and we have already encoded the
expected temperature dependence of the quadratic terms close to the onset of non-zero
order parameters. 

The minimization of the free energy $F$ determines thermal equilibrium, whose stability requires $b$, $b' > 0$. Depending on its coefficients, one finds in general four possible
solutions to the minimization of $F$: a solution where both order parameters vanish, 
two solutions where one of the order parameter is vanishing, 
and a solution with a coexistence of both order parameters. 
The observed charge-order in the absence of magnetization
implies that $T_{\rm C'} < T_{\rm P}$, and $a$, $a' > 0$, so that $\phi$ orders at $T=T_{\rm P}$.
When the coupling constant $\gamma$ satisfies the constraints $ab'/a' < \gamma < a'b/a$ and $\gamma^2<bb'$,
$\phi$ exhibits a maximum
at the magnetic critical temperature $T_{\rm C} = (a'bT_{\rm C'}-\gamma aT_{\rm P}) / (a'b-\gamma a)$. 

\begin{figure}
	\begin{minipage}[t]{0.55\textwidth}
	\includegraphics[width=0.99\columnwidth]{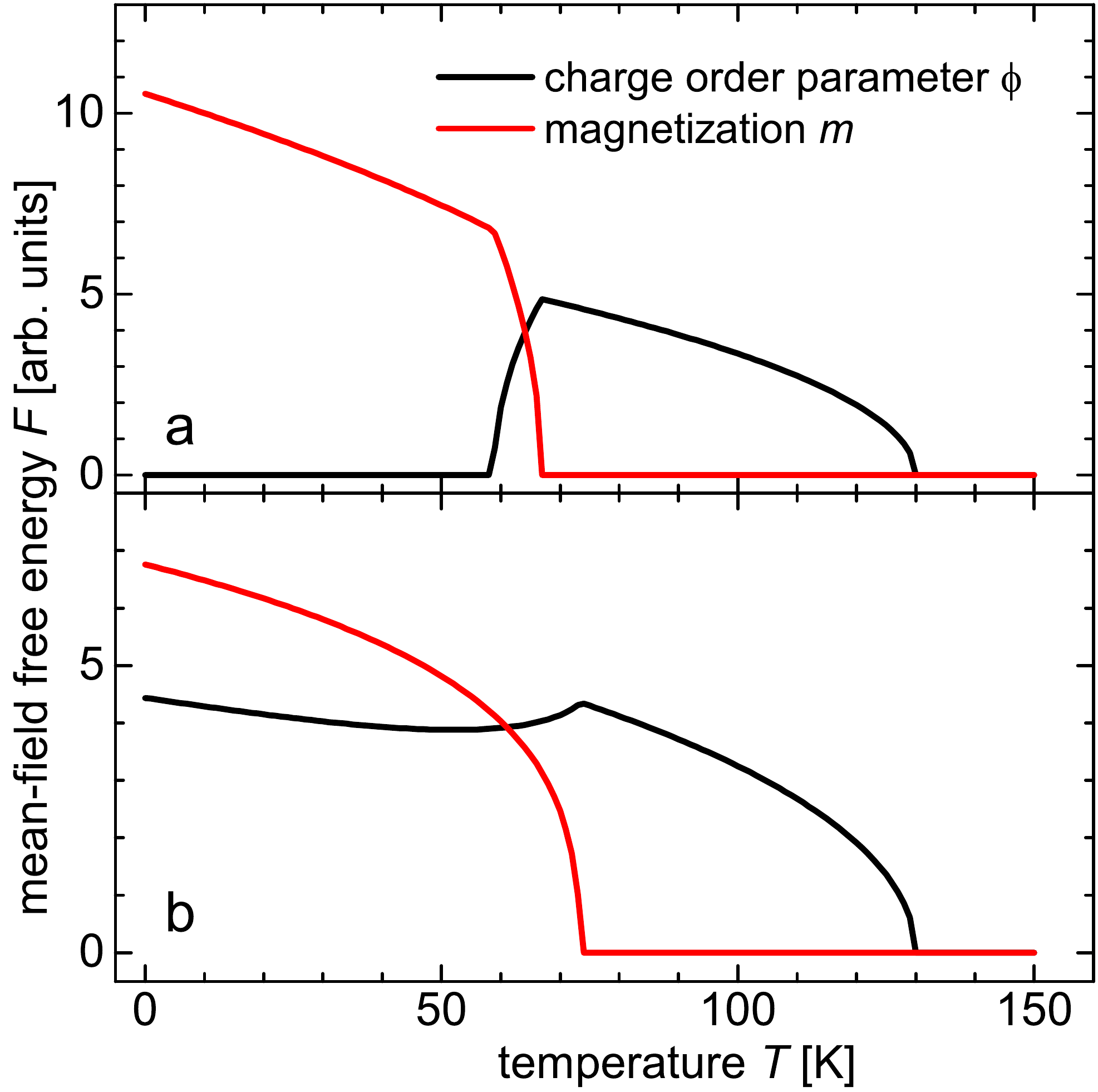}
	\end{minipage}
	\hfill
	\begin{minipage}[b]{0.41\textwidth}
	\caption{{\bf Expansion of the temperature-dependent Ginzburg-Landau free energy.} 
	Plots showing the results of an expansion of the GL free energy $F$ to the forth (({\bf a}) parameters: 
	$a = 0.9$, $a' = 1$, $T_{\rm P} = 130$, $T_{\rm C'} = 100$, $b = 2.4$, $b' = 0.9$, $\gamma = 1.4$, $c = c' = 0$), 
	and to the sixth power (({\bf b}) same parameters except for $c = c' = 0.015$), 
	as described in Eqs.~(\ref{LandauF}) and (\ref{LandauF6}), respectively.}
	\label{phim}  \end{minipage}   
\end{figure}

An example of the resulting order parameters is shown in Fig.~\ref{phim}a. 
We observe that this solution displays an interval of temperature 
where $\phi$ vanishes while $m > 0$. 
This behavior is a result of the competition between the $\frac{a}{2}\left(T-T_{\rm P}\right)\phi^2$ term, 
which is negative for $T<T_{\rm P}$, and the positive coupling with $m$, $\frac{\gamma}{2}\phi^2m^2$: 
upon decreasing the temperature, $m$ grows and a strong enough coupling $\gamma$ 
pushes down the value of $\phi$ which minimizes $F$, eventually leading to $\phi = 0$.
However, this zero of the charge-order parameter is unstable 
with respect to the inclusion of higher-order terms in Eq.~(\ref{LandauF}).
Although such corrections are irrelevant close to the phase transitions of
$\phi$ and $m$, they nevertheless influence their growth in a wider range of temperatures.
In fact, the expansion of Eq.~(\ref{LandauF}) up to the fourth order 
predicts the order parameters to grow indefinitely below the critical temperature, whereas
in real materials a saturation effect is expected.
This suggests to study an improved GL free energy expansion, including next-to-leading powers:

\begin{multline}
	F = F_0 + \frac{a}{2}\left(T-T_{\rm P}\right)\phi^2 + \frac{b}{4}\phi^4 + \frac{c}{6}\phi^6
	+ \frac{a'}{2}\left(T-T_{\rm C'}\right)m^2+\frac{b'}4m^4+\frac{\gamma}{2}\phi^2m^2+\frac{c'}{6}m^6,
\label{LandauF6}
\end{multline}
where stability requires $c$, $c' > 0$.
The inclusion of the sixth power qualitatively changes the behavior of $\phi$,
giving rise to a dip
qualitatively similar to the experimentally observed behavior. 
In other words, the solution with a vanishing charge-order 
and non-zero magnetization requires a fine-tuning of the higher-order terms $c = c' = 0$, 
whereas their inclusion naturally explains the observed dip in $\phi$. 
The sixth term effectively damps the growth of $m$ away from $T=T_{\rm C}$, 
such that $\phi$ is not pushed down to $0$, but instead displays a minimum.

\section{Discussion} 
Our experiments show that a sample system that is conceptually as simple as 
the pseudomorphic Fe double-layer of Fe/Rh(001) may possess different ordering phenomena.  
Even more importantly, our data reveal that the ordering phenomena at play here, i.e.\ charge-order and ferromagnetism, 
compete with each other as evidenced by the intermediate reduction of the charge-order parameter $\phi$ 
observed experimentally and in GL calculations. 
We speculate that this cross-talk is caused by the fact that the electronic structure close to the Fermi level, 
which is responsible for charge- and magnetic-order phenomena 
through Fermi surface nesting and the exchange interaction, respectively,
drastically changes at the respective critical temperatures.  
We expect that details of the temperature-dependent evolution of the electronic structure 
will be subject of future experiments, such as angular-resolved photoemission spectroscopy, 
to shed light on the subtle balance between nested and exchange-split electronic states.

\section{Methods}
{\bf STM, STS, and SPSTM measurements.} Scanning tunneling microscopy (STM) measurements 
were performed under ultrahigh vacuum ($p \leq 5.0 \times 10^{-11}$\,mbar) 
with a home-built low-temperature (LT)-STM and a commercial variable-temperature (VT)-STM 
at sample temperatures of $T_{\rm LT} = 5$\,K and $T_{\rm VT} = 30 ... 300$ K, respectively. 
STM tips were prepared from electro-chemically etched tungsten (W) wires 
which were flashed under ultrahigh vacuum conditions. 
For spin-resolved measurements  the W tips were coated with $\approx 20$ atomic layers (AL) 
of Cr, resulting in out-of-plane magnetic sensitivity, as verified by test measurements 
on samples with well-known magnetization directions, i.e.\ Fe/W(110) or Mn/W(110) \cite{MBode1}. 
For scanning tunneling spectroscopy (STS) measurements a small modulation 
was added to the sample bias voltage $U$ (frequency $\nu = 5.777$\,kHz; amplitude 5 to 15\,mV), 
such that tunneling differential conductance d$I/$d$U$ spectra as well as d$I/$d$U$ maps 
can be acquired by detecting the first harmonic signal with a lock-in amplifier. 
 
{\bf Sample preparation of Fe/Rh(001).} The Rh(001) surface was prepared by cycles that consist of 
about 10\,min Ar ion sputtering at room temperature 
($p_{\rm Ar} = 5 \times 10^{-6}$\,mbar, $E_{\rm ion} = 1$\,keV), 
followed by 150\,sec annealing at $T_{\rm an} = 1300$\,K in an oxygen atmosphere, 
and a final flash (duration about 30\,sec) without oxygen at the same temperature.  
It has been shown that this procedure reliably removes carbon impurities from the surface \cite{KWK2015}. 
Subsequently, Fe films were grown on the Rh(001) surface at $T = 315$\,K by means of the e-beam evaporation. 


\section{Acknowledgments}
This work has been funded by Deutsche Forschungsgemeinschaft 
within BO 1468/22-1 and through SFB 1170 "ToCoTronics". 
\section{Author contributions}
P.-J.H., J.K., J.K., T.B., and M.V. performed the STM experiment and jointly with M.B. analyzed the experimental data.
F.P.T. and F.A. performed Ginzburg-Landau modeling.   All authors contributed to the scientific discussion.
P.-J.H., F.P.T. and M.B. wrote the manuscript with input and comments from all co-authors.
\section{Additional information}
Supplementary information is included in this submission.
\section{Competing financial interests}
The authors declare no competing financial interests.

\bibliography{article}

\begin{references}


\bibitem{JHoffman} Hoffman, J. E. \textit{et al.} 
A four unit cell periodic pattern of quasi-particle states surrounding vortex cores 
in Bi$_{2}$Sr$_{2}$CaCu$_{2}$O$_{8+\delta}$. 
\textit{Science} {\bf 295}, 466-469 (2002).

\bibitem{JDavis} Davis, J. C. S. $\&$ Lee, D. H. 
Concepts relating magnetic interactions, intertwined electronic orders, 
and strongly correlated superconductivity. 
\textit{Proc. Natl. Acad. Sci. U.S.A.} {\bf 110}, 17623-17630 (2013). 

\bibitem{PCai} Cai, P. \textit{et al.} 
Visualizing the microscopic coexistence of spin density wave and superconductivity 
in underdoped NaFe$_{1-x}$Co$_{x}$As. 
\textit{Nat. Commun.} {\bf 4}, 1596 (2013). 

\bibitem{TKimura} Kimura, T. \textit{et al.} 
Magnetic control of ferroelectric polarization. 
\textit{Nature} {\bf 426}, 55-58 (2003).

\bibitem{MKenzelmann} Kenzelmann, M. \textit{et al.} 
Magnetic inversion symmetry breaking and ferroelectricity in TbMnO$_{3}$. 
\textit{Phys. Rev. Lett.} {\bf 95}, 087206 (2005).

\bibitem{SCheong} Cheong, S. W. $\&$ Mostovoy, M. 
Multiferroics: a magnetic twist for ferroelectricity. 
\textit{Nature Mat.} {\bf 6}, 13-20 (2007).

\bibitem{PRL85_3720} Garc\'{i}a, D. J., Hallberg, K., Batista, C. D., Avignon, M. and Alascio, B. 
New Type of Charge and Magnetic Order in the Ferromagnetic Kondo Lattice. 
\textit{Phys. Rev. Lett.} {\bf 85}, 3720 (2000).

\bibitem{DADikin} Dikin, D. A. \textit{et al.} 
Coexistence of superconductivity and ferromagnetism in two dimensions. 
\textit{Phys. Rev. Lett.} {\bf 107}, 056802 (2011).

\bibitem{LLi} Li, L. \textit{et al.} 
Coexistence of magnetic order and two-dimensional superconductivity at LaAlO$_{3}$/SrTiO$_{3}$ interfaces. 
\textit{Nature Phys.} {\bf 7}, 762-766 (2011).

\bibitem{JABert} Bert, J. A. \textit{et al.} 
Direct imaging of the coexistence of ferromagnetism and superconductivity at the LaAlO$_{3}$/SrTiO$_{3}$ interface. 
\textit{Nature Phys.} {\bf 7}, 767-771 (2011).

\bibitem{Sachdev_Science} Sachdev, S. 
Quantum Criticality: Competing Ground States in Low Dimensions. 
\textit{Science} {\bf 288}, 475-480 (2000).

\bibitem{Kittel} Kittel, C. \textit{Introduction to Solid State Physics} 6th edn, (John Wiley $\&$ Sons. 1986).

\bibitem{CSWang} Wang, C. S. \textit{et al.} 
Theory of magnetic and structural ordering in iron. 
\textit{Phys. Rev. Lett.} {\bf 54}, 1852 (1985).

\bibitem{FJPinski} Pinski, F. J. \textit{et al.} 
Ferromagnetism versus antiferromagnetism in face-centered-cubic iron.
\textit{Phys. Rev. Lett.} {\bf 56}, 2096 (1986).

\bibitem{TAsada} Asada, T. \textit{et al.} 
Total energy spectra of complete sets of magnetic states for fcc-Fe films on Cu(100). 
\textit{Phys. Rev. Lett.} {\bf 79}, 507 (1997).

\bibitem{DSpisak} Spi\u{s}\'{a}k, D. \textit{et al.} 
Spin-density wave in ultrathin Fe films on Cu(100). 
\textit{Phys. Rev. B} {\bf 66}, 052417 (2002).

\bibitem{KKnopfle} Kn{\"o}pfle, K. \textit{et al.} 
Spin spiral ground state of $\gamma$-iron. 
\textit{Phys. Rev. B} {\bf 62}, 5564 (2000).

\bibitem{SHeinze} Heinze, S. \textit{et al.} 
Spontaneous atomic-scale magnetic skyrmion lattice in two dimensions. 
\textit{Nature Phys.} {\bf 7}, 713-718 (2011).

\bibitem{DQian} Qian, D. \textit{et al.} 
Spin-density wave in ultrathin Fe films on Cu(100). 
\textit{Phys. Rev. Lett.} {\bf 87}, 227204 (2001).

\bibitem{AKubetzka} Kubetzka, A. \textit{et al.} 
Revealing antiferromagnetic order of the Fe monolayer on W(001): 
spin-polarized scannning tunneling microscopy and first-principles calculations. 
\textit{Phys. Rev. Lett.} {\bf 94}, 087204 (2005).

\bibitem{SMeckler} Meckler, S. \textit{et al.} 
Real-space observation of a right-rotating inhomogenous cycloidal spin spiral 
by spin-polarized scanning tunneling microscopy in a triple axes vector magnet. 
\textit{Phys. Rev. Lett.} {\bf 103}, 157201 (2009).

\bibitem{NRomming} Romming, N. \textit{et al.} 
Writing and deleting single magnetic skyrmions. 
\textit{Science} {\bf 341}, 636-639 (2013).

\bibitem{PhysRevB.64.054417}
Hayashi, K., Sawada, M., Harasawa, A., Kimura, A. \& Kakizaki, A.
Structure and magnetism of Fe thin films grown on Rh(001) studied by photoelectron spectroscopy.
\textit{Phys. Rev. B} {\bf 64}, 054417 (2001).

\bibitem{Hayashi_JPSJap}
Hayashi, K., Sawada, M., Yamagami, H., Kimura, A. \& Kakizaki, A.
Magnetic Dead Layers Induced by Strain at fat Fe/Rh(001) Interface.
\textit{J. Phys. Soc. Jpn.} {\bf 73}, 2550-2553 (2004).

\bibitem{DSpisak2} Spi\u{s}\'{a}k, D. \textit{et al.} 
Structural, magnetic and chemical properties of thin Fe films grown on Rh(100) surfaces 
investigated with density functional theory. 
\textit{Phys. Rev. B} {\bf 73}, 155428 (2006).

\bibitem{AlZubi} Al-Zubi, A. \textit{et al.} 
Magnetism of 3$\textit{d}$ transition-metal monolayers on Rh(100). 
\textit{Phys. Rev. B} {\bf 83}, 024407 (2011).

\bibitem{MTakada} Takada, M. \textit{et al.} 
A complex magnetic spin structure of ultrathin Fe films on Rh(001) surfaces. 
\textit{J. Mag. Mag. Mat.} {\bf 329}, 95-100 (2013).

\bibitem{KWK2015}
Kemmer, J., Wilfert, S., K{\"u}gel, J., Mauerer, Hsu, P.-J., \& Bode, M. 
Growth and magnetic domain structure of ultra-thin Fe-films on Rh(001).
\textit{Phys. Rev. B} {\bf 91}, 184412 (2015).


\bibitem{MBode1} Bode, M. 
Spin-polarized scanning tunneling microscopy. 
\textit{Rep. Prog. Phys.} {\bf 66}, 523-582 (2003).

\bibitem{Claessen1990} Claessen, R., Burandt, B., Carstensen, H. \& Skibowski, M. 
Conduction-band structure and charge-density waves in 1T-TaS$_2$.
\textit{Phys. Rev. B} {\bf 41}, 8270 (1990).

\bibitem{Rossnagel2002} Rossnagel, K., Kipp, L. \& Skibowski, M. 
Charge-density-wave phase transition in 1T-TiSe$_2$: 
Excitonic insulator versus band-type Jahn-Teller mechanism. 
\textit{Phys. Rev. B} {\bf 65}, 235101 (2002).

\bibitem{Valla2000} Valla, T., Fedorov, A.V.,  Johnson, P.D., Xue, J.K., Smith, E. \& DiSalvo, F.J. 
harge-Density-Wave-Induced Modifications to the Quasiparticle Self-Energy in 2H-TaSe$_2$. 
\textit{Phys. Rev. Lett.} {\bf 85}, 4759 (2000).

\bibitem{critbook}
Nishimori, H. \& Ortiz, G.
Elements of Phase Transitions and Critical Phenomena.
\textit{Oxford Graduate Texts}, OUP Oxford, ISBN 9780191035531 (2010).

\end{references}
\bibliographystyle{apsrevkuo}

\end{document}